# Implementing Software Defined Load Balancer and Firewall

Shreya Rajkumar*

*(ECE, Georgia Instituteof Technology, United States

Email: shreyarajkumar2@gmail.com)

----------------------------------------************************----------------------------------

## Abstract:

Software-defined networking (SDN) is an architecture that aims to make networks fast and flexible. SDN's goal is to improve network control by enabling service providers as well as enterprises to respond quickly to changing business needs. In SDN, the administrator can shape traffic from a centralized control console without having to modify any of the individual switches belonging to the network. The SDN controller which is centralized directs the switches to deliver network services wherever they are needed, irrespective of the specific connections between a server and devices. This methodology is a shift from traditional network architecture, in which individual network devices make traffic decisions based on their configured routing tables. In this paper, I built and tested an SDN load balancer and firewall module using the Floodlight controller.

----------------------------------------************************----------------------------------

## I. INTRODUCTION

SDN has been around for a while now and shows a lot of potential in making things easier when it comes to dynamic things in the network. Firewall and load balancers are by nature dynamic and often changefrequently in any network setup. Considering this, Load balancers and Firewalls seem to be an exact fit for the kind of innovations SDN can bring in and can be most beneficial in.

Some of the points making a strong case for SDN Load balancers and Firewalls are:

1) Cloud-Native Applications: With the world moving on to the cloud through VMs and containers, a remotely controlled software component is easier to handle than having hardware components that are difficult to maintain when the data center is thousands of miles away from the users.

2) Scalability: To scale an SDN component can be as easy as adding a new controller to the system which requires minimum overhead and is equivalent to deploying another software component on the cloud.

3) Flexibility: Insoftware-defined load balancing and firewalls, administrators can deploy custom application services on a per-application basis, instead of fitting multiple applications on a single monolithic hardware appliance to save hardware costs.

4) Hybrid cloud applications: Software load balancers and firewalls provide a consistent application delivery architecture across different cloud environments.

5) Thiseliminates the need to re-architect applications when migrating to the cloud or between clouds.super convenient and any new updates/fixes can be rolled out seamlessly to all the controllers without much hassle and risk.

6) Redundancy/Resilience: If a server running the load balancer/firewall is brought down, other deployments can be quickly enabled to pick up the slack and prevent service disruptions.

## II. OVERVIEW

A Load balancer and a Firewall were built using SDN. The floodlight controller was leveraged, and python modules were written to interact with this controller. The network was simulated using Mininet as seen in Fig 1, and the results were analyzed using Wireshark. Section III details the methodology. Section IV lists related work carried out in the SDN sphere, and section V concludes the work with thoughts on future work.

Fig.1 Mininet Topology

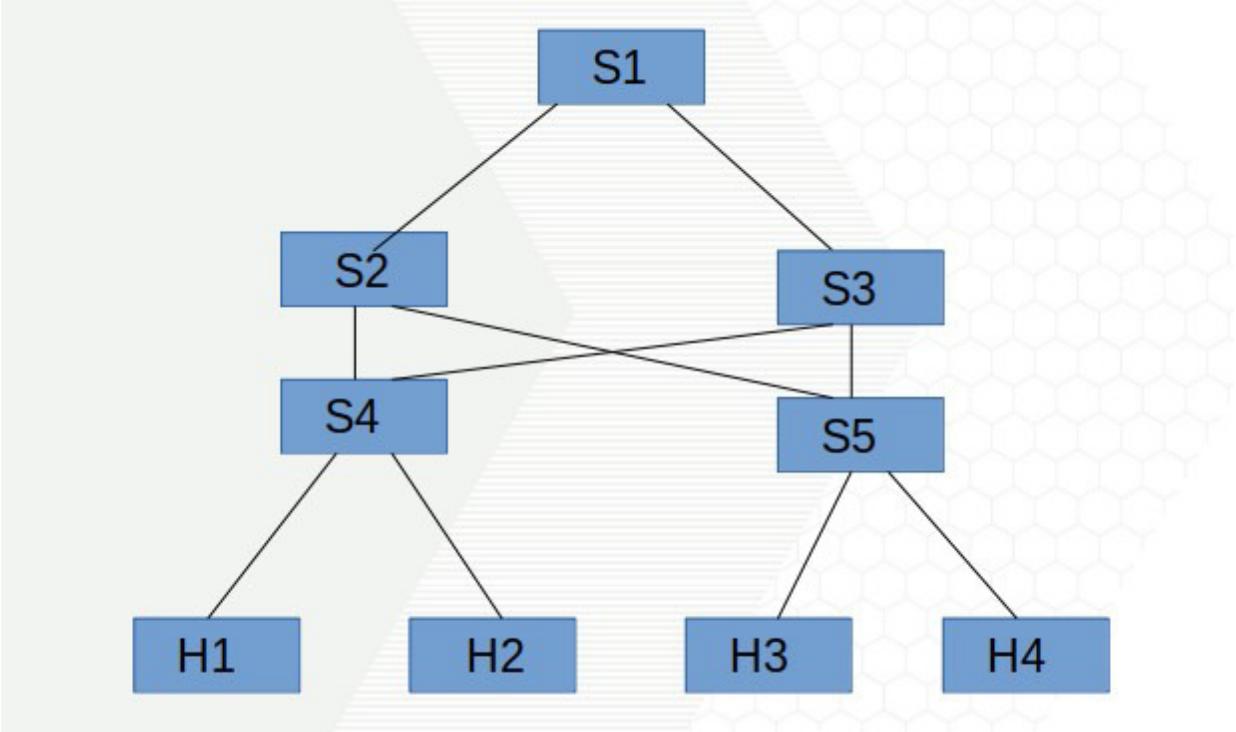

## III. METHODOLOGY

To start with the implementation of the Load Balancer, I first set up Mininet and Floodlight on the systems. A realistic virtual network is created by Mininet, running real kernel, switch, and application code, on a single machine [6]. For any SDN-based study, the most important thing is the choice of the controller. After thorough research on the various SDN controllers available and weighing their pros and cons, I decided to choose the Floodlight controller. The Floodlight Open SDN Controller is an enterprise-class, Apache-licensed, Java-based OpenFlow Controller [5]. The main reasons for choosing this controller were because of the clear documentation as well as the presence of REST APIs to simplify application interfaces to the product.

### A. *MininetTopology*

I sought to simulate my project in a rudimentary 3-tier topology which is displayed in Fig. 1. H1-H4 are leaf nodes, while S1-S5 are switches, which will route packets according to the policies that were pushed by the flood light controller.

### B. *LoadBalancer*

After a good understanding of the working of the entire setup, I decided to implement a very simple load balancing module to start with. The flow diagram for the same is displayed in Fig. 2. The python module will interact with floodlight, which in turn will push rules to the desired switches. For this, I implemented a static rule-based load balancer which works as follows:

1) The module uses Floodlight REST APIs to push static loadbalancingrulesinformationtoFloodlight.
2) The rule contains a static (hardcoded) destination IP address and the link to forward the traffic on. This way I was able to achieve two different packets with different destination IP being forwarded to different interfacesaspertherulespushedbyus.
3) Once the rules were loaded into the controller, I verified the correct functioning of the module through packet tracing usingWireshark.

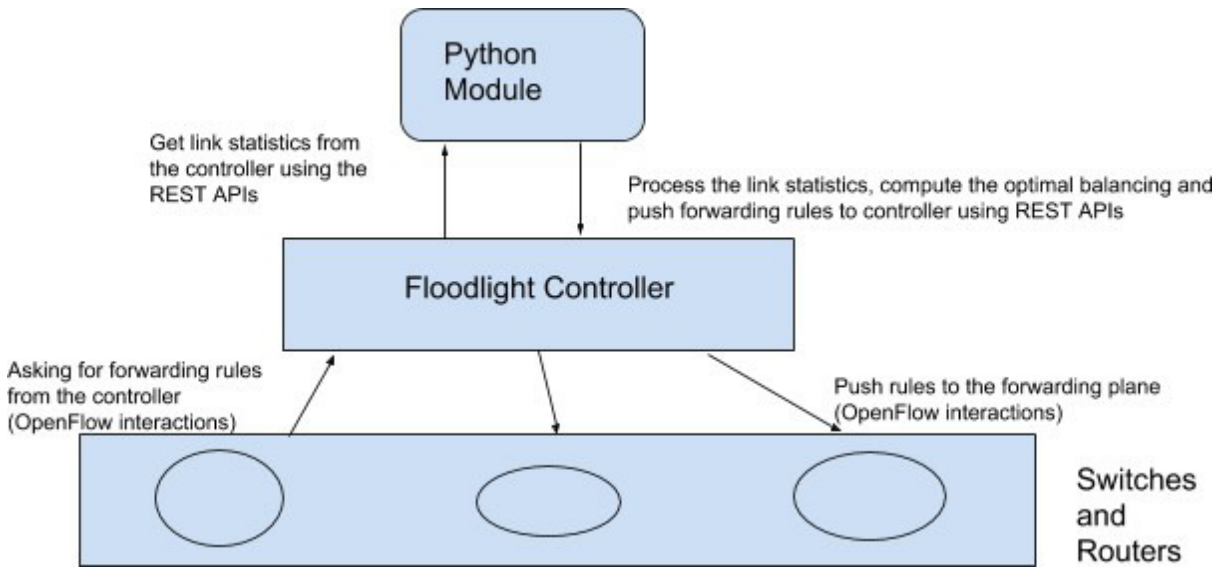

Fig.2 Interactions among components in the SDN load balancer

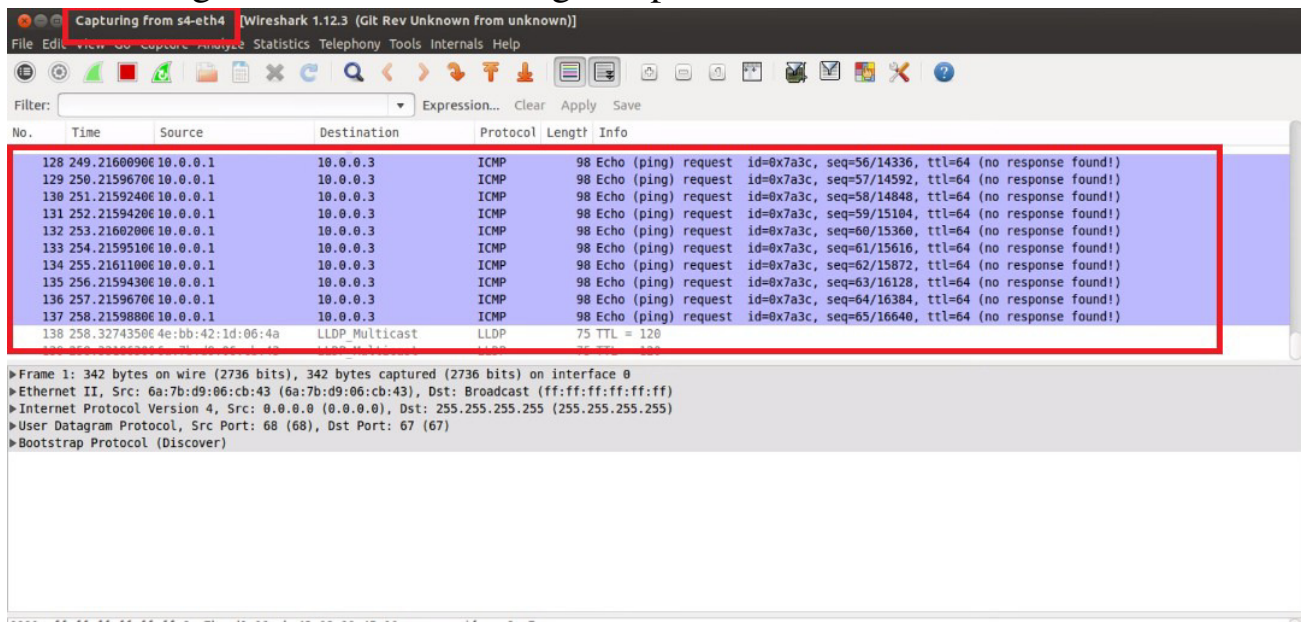

balancer module implements the following algorithm:

1) Identify all paths from the source to the destination in the form of a list of links.
2) Fetch the transmission rate (Tx bits per sec) for all the links in all the paths.
3) For every path, the cost is defined as the sum of the transmission rate across all the links in the path.
4) Thecost forevery identified path is calculated.
5) The path with the minimum cost is chosen and the corresponding flow is pushed to the controller.

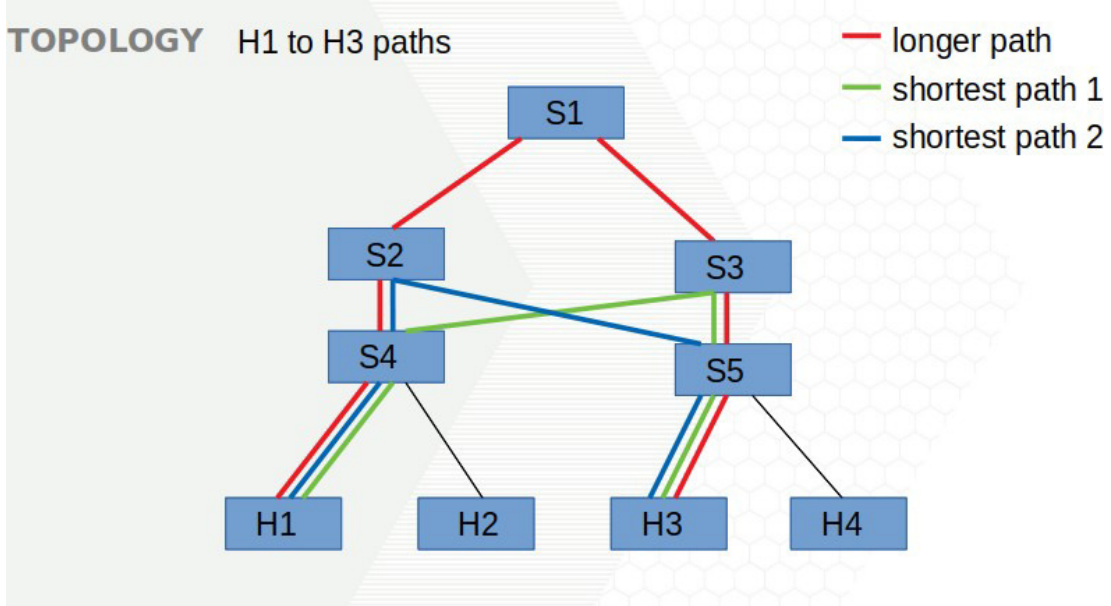

Fig.3 Shortest and longest paths in the topology from H1 to H4

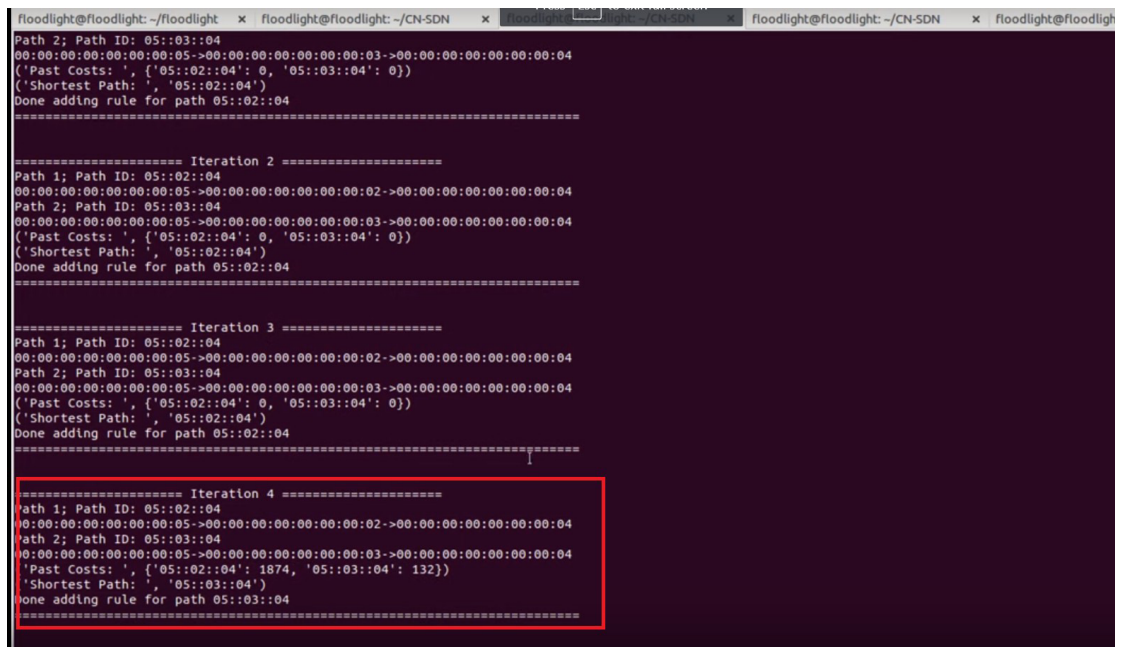

Fig.4 Identification of all the paths and the costs by the load balancer

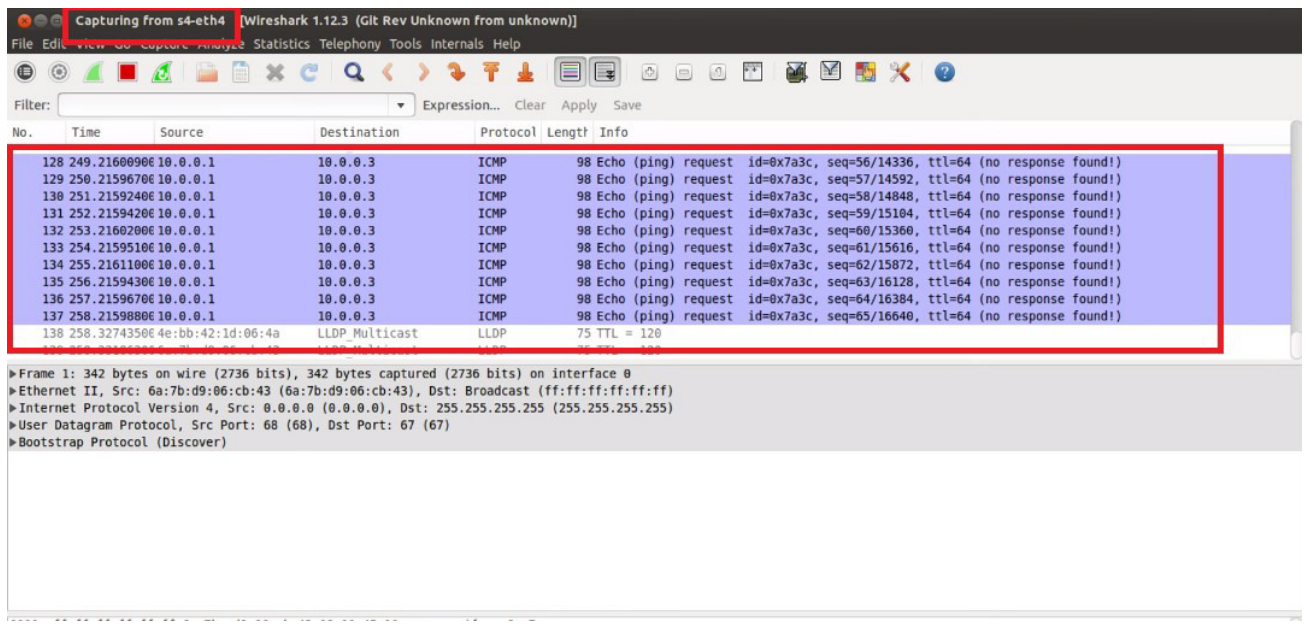

Fig.5 WiresharkshowingtrafficfromH1toH3takingblueroute

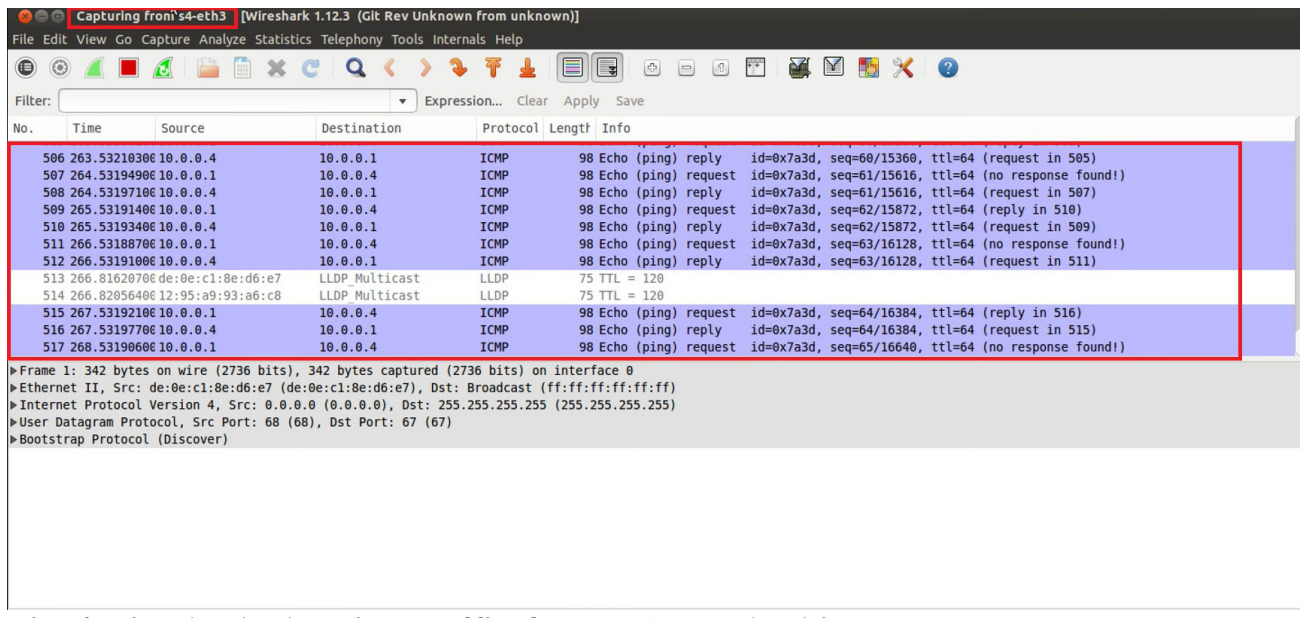

Fig.6Wireshark showing traffic from H1 to H4 taking green route

Once I had the model ready, I simulated ping traffic across all nodes and explicitly tried to ping H1-H3, and H1-H4. I then observed the traffic flow on Wireshark to notice that before load balancing, both flows opted for H1-S4- S2-S5, but soon after a few iterations of the load balancing module for balancing load between H1 and H4, I found the flow had switched to H1-S4-S3-S5 path.

The figures 4, 5, and 6 illustrate the identification of the paths and the costs, and Wireshark confirms balanced load between flows H1 to H3 and H1 to H4 respectively.

## C. Firewall

The Firewall module is simpler in approach. The rules are applied as per user specification and pervade the entire network. The module takes in user parameters takes in IP, port, and operation, and then flushes the rule to the entire network via the controller. If a packet arrives at a switch that matches the rule flushed, appropriate action is taken on the packet.

Firstly, the Floodlight firewall is disabled. The user can then specify the source and destination IP address and whether he would like to enable or disable a particular flow on the go. The IP addresses mentioned will be added to the flow and the corresponding action will be taken (allow or deny).

Figures 7, 8, 9, and 10 depict the working of the firewall.

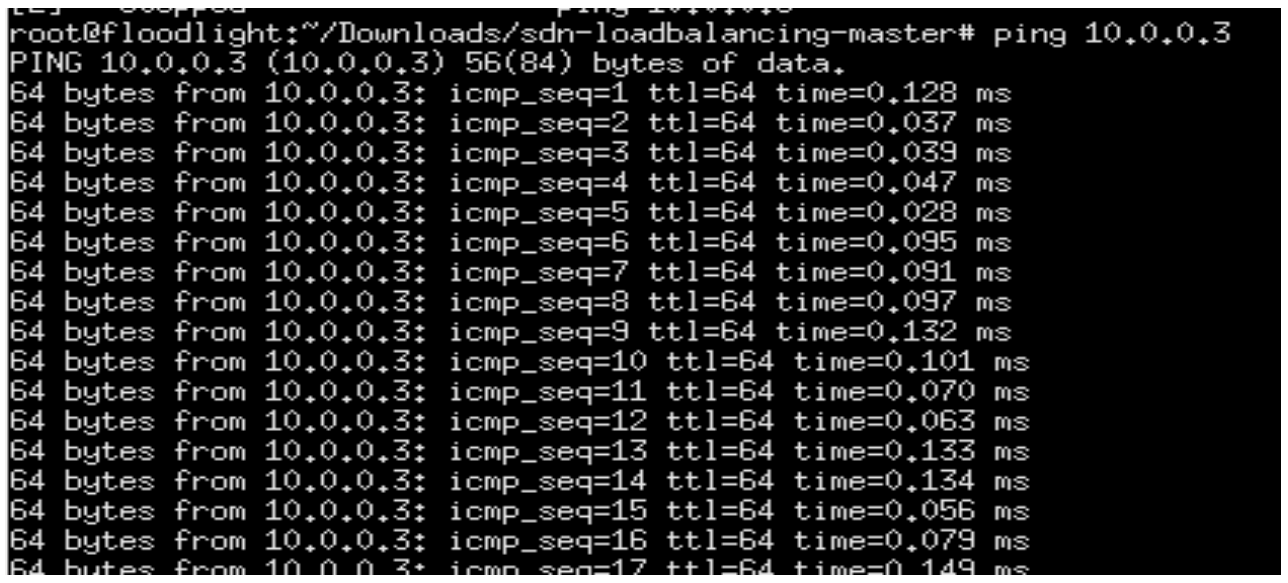

Fig.7 Packets sent from source to destination.

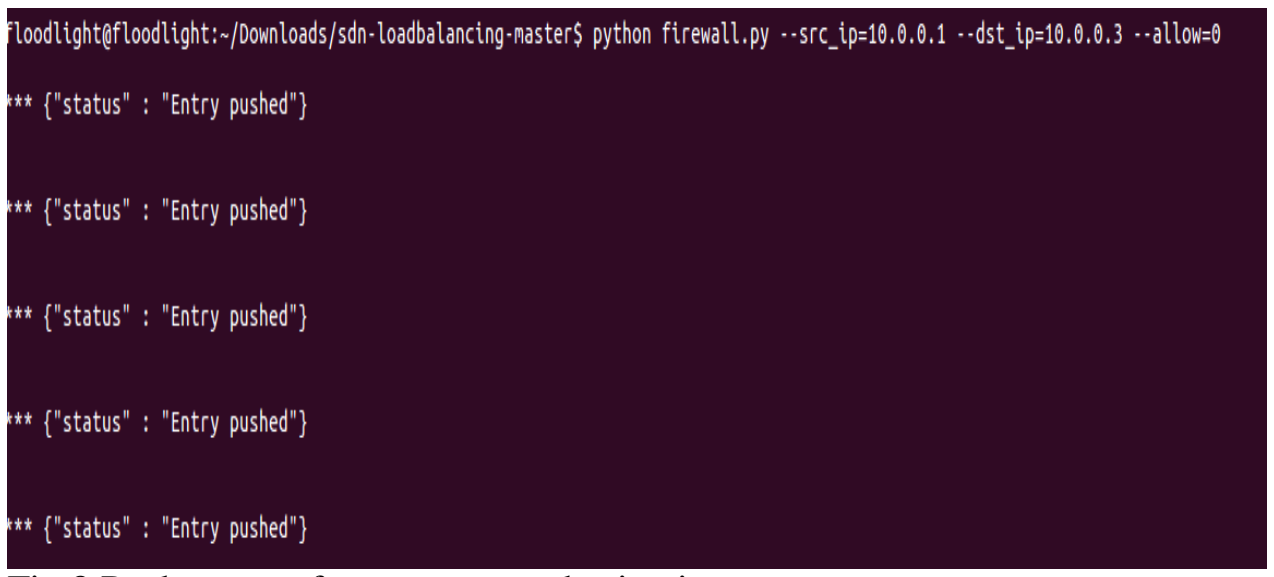

Fig.8 Packets sent from source to destination.

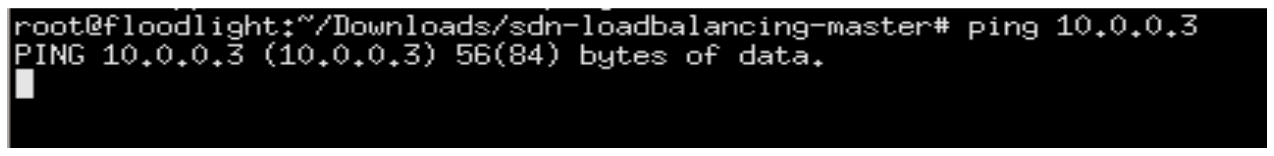

Fig.9No packets sent from source to destination

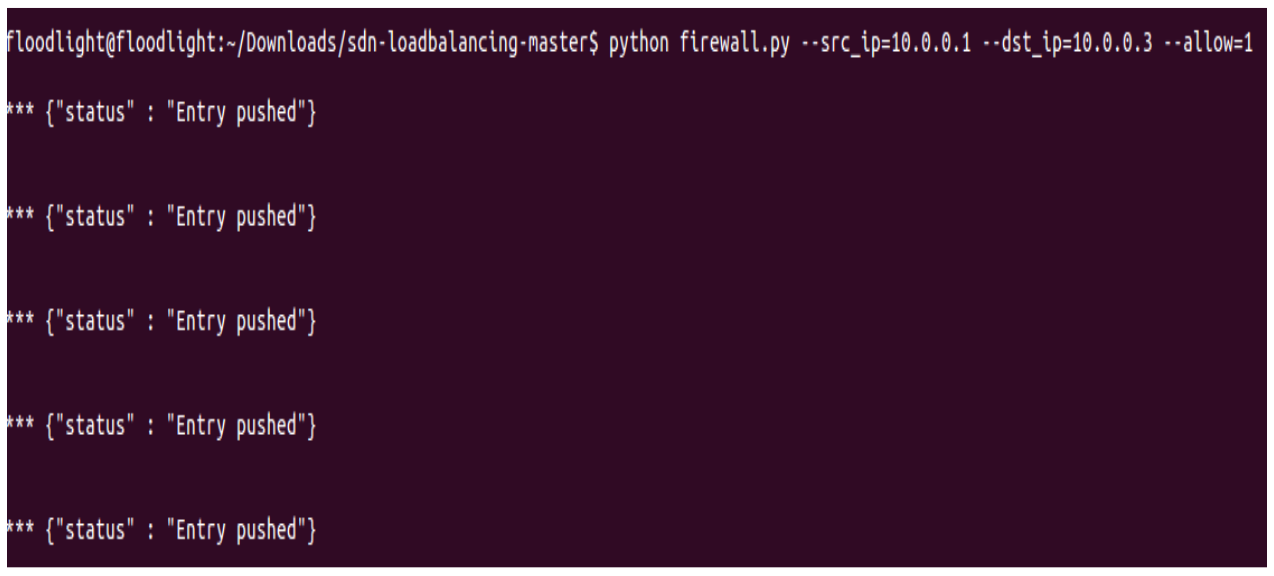

Fig.10 User enables flow by setting allow=1

When the user mentions allowing as 1, the flow is enabled from that source to the destination. Similarly, when the user mentions allow as 0, the flow is disabled from that source to destination and the switches along that path are blocked.

## IV. RELATED WORK

I primarily surveyed literature regarding load balancing algorithms and firewalls. The several types of load balancing algorithms can broadly be classified into static and dynamic. A few are described below:

1) ROUND-ROBIN: In this, each server receives the request from clients in a circular manner. The requests are allocated to various live servers on a round-robin base.[1]

2) WEIGHTED ROUND-ROBIN: In this, each server receives the request from the client based on criteria that are fixed by the site administrator. In other words, a static weight is assigned to each server in Weighted Round Robin (WRR) policy. We usually specify weights in proportion to actual capacities. So, for example, if Server 1's capacity is 5 times more than Server 2's, then we can assign a weight of 5 to server 1 and a weight of 1 to server 2.[1]

3) RANDOM STRATEGY: From a list of live servers, the load balancer will randomly choose a server for sending requests. This policy has large overheads.[1]

4) HASH-BASED. The algorithm works by first calculating the hash value of traffic flow using the IP address of source and destination, port number of source and destination, and URL. The request is then forwarded to the server with the highest hash value. If any other request comes with the same hash value, it will be forwarded to the same server.[2]

5) GLOBAL FIRST FIT(GFF): After receiving a new flow request, the scheduler searches linearly all the available paths to find the one which can accommodate the bandwidth requirement of this new flow. The flow is greedily assigned to the first path which is fulfilling the requirement. Global First Fit does not distribute flows evenly across all paths.[3]

6) FLOW-BASED LOAD BALANCING: Classify the in- coming request into mice or elephant flows. The serverselection module will select the server with the smallest elephant flow counter value or mice flow counter value depending on whether the new flow is elephant or mice, respectively. Collect port statistics and flow statistics of switches. The statistics help in the selection of Server/Path based on a weighted heuristic.[3]

Firewalls have been implemented using simple rules based on MAC, IP addresses, protocol, etc. [4].

## V. RESULTS

The load balancer module and firewall were successfully implemented and tested with varying topologies and environments. The exact experiments and screenshots are explained in the Methodology part of the paper.

The possible enhancements and scope of improvements in both modules are discussed further in Future Work.

## VI. FUTURE WORK

My current load balancing algorithm is based on transmission rate (bits per second through links). This might be a very limiting heuristic, as we would want certain traffic to flow via specific paths irrespective of load on the path. Future work on the same would see implementations of the algorithms that are based on other link characteristics such as rate of loss, RTT, etc. Policies can also factor in priority and the content of the traffic. A beneficial feature for the end-user of the module would be flexibility in selecting an algorithm.

The present firewall operates on user input and is limited to IP and port. This implementation can be extended to take actions based on packet contents or to specify access lists that depend on the source of originating traffic.

## ACKNOWLEDGMENT

I would like to thank Georgia Institute of Technology for giving me the opportunity to research on this topic.